\documentstyle[12pt]{article}
\def\PRL #1 #2 #3{{\sl Phys. Rev. Lett.} {\bf#1} (#2) #3}
\def\NPB #1 #2 #3{{\sl Nucl. Phys.} {\bf B #1} (#2) #3}
\def\NPBFS #1 #2 #3 #4{{\sl Nucl. Phys.} {\bf B #2} [FS#1] (#3) #4}
\def\CMP #1 #2 #3{{\sl Commun. Math. Phys.} {\bf #1} (#2) #3}
\def\PRD #1 #2 #3{{\sl Phys. Rev.} {\bf D #1} (#2) #3}
\def\PLA #1 #2 #3{{\sl Phys. Lett.} {\bf #1A} (#2) #3}
\def\PLB #1 #2 #3{{\sl Phys. Lett.} {\bf B #1} (#2) #3}
\def\JMP #1 #2 #3{{\sl J. Math. Phys.} {\bf #1} (#2) #3}
\def\PTP #1 #2 #3{{\sl Prog. Theor. Phys.} {\bf #1} (#2) #3}
\def\SPTP #1 #2 #3{{\sl Suppl. Prog. Theor. Phys.} {\bf #1} (#2) #3}
\def\AoP #1 #2 #3{{\sl Ann. of Phys.} {\bf #1} (#2) #3}
\def\PNAS #1 #2 #3{{\sl Proc. Natl. Acad. Sci. USA} {\bf #1} (#2) #3}
\def\RMP #1 #2 #3{{\sl Rev. Mod. Phys.} {\bf #1} (#2) #3}
\def\PR #1 #2 #3{{\sl Phys. Reports} {\bf #1} (#2) #3}
\def\AoM #1 #2 #3{{\sl Ann. of Math.} {\bf #1} (#2) #3}
\def\UMN #1 #2 #3{{\sl Usp. Mat. Nauk} {\bf #1} (#2) #3}
\def\FAP #1 #2 #3{{\sl Funkt. Anal. Prilozheniya} {\bf #1} (#2) #3}
\def\FAaIA #1 #2 #3{{\sl Functional Analysis and Its Application} {\bf
#1} (#2) #3}
\def\BAMS #1 #2 #3{{\sl Bull. Am. Math. Soc.} {\bf #1} (#2)
#3} \def\TAMS #1 #2 #3{{\sl Trans. Am. Math. Soc.} {\bf #1} (#2) #3}
\def\InvM #1 #2 #3{{\sl Invent. Math.} {\bf #1} (#2) #3}
\def\LMP #1 #2 #3{{\sl Letters in Math. Phys.} {\bf #1} (#2) #3}
\def\IJMPA #1 #2 #3{{\sl Int. J. Mod. Phys.} {\bf A #1} (#2) #3}
\def\AdM #1 #2 #3{{\sl Advances in Math.} {\bf #1} (#2) #3}
\def\RMaP #1 #2 #3{{\sl Reports on Math. Phys.} {\bf #1} (#2) #3}
\def\IJM #1 #2 #3{{\sl Ill. J. Math.} {\bf #1} (#2) #3}
\def\APP #1 #2 #3{{\sl Acta Phys. Polon.} {\bf #1} (#2) #3}
\def\TMP #1 #2 #3{{\sl Theor. Mat. Phys.} {\bf #1} (#2) #3}
\def\JPA #1 #2 #3{{\sl J. Physics} {\bf A#1} (#2) #3}
\def\JSM #1 #2 #3{{\sl J. Soviet Math.} {\bf #1} (#2) #3}
\def\MPLA #1 #2 #3{{\sl Mod. Phys. Lett.} {\bf A #1} (#2) #3}
\def\JETP #1 #2 #3{{\sl Sov. Phys. JETP} {\bf #1} (#2) #3}
\def\JETPL #1 #2 #3{{\sl  Sov. Phys. JETP Lett.} {\bf #1} (#2) #3}
\def\PHSA #1 #2 #3{{\sl Physica} {\bf A #1} (#2) #3}
\def\CQG #1 #2 #3{{\sl Class. Quantum Grav.} {\bf #1} (#2) #3}
\def\SJNP #1 #2 #3{{\sl Sov. J. Nucl. Phys. (Yadern.Fiz.)} {\bf #1} (#2) #3}
\def\a{\alpha}\def\b{\beta}\def\g{\gamma}\def\d{\delta}

\def\l{\lambda}\def\s{\sigma}
\def\Om{\Omega}

\newcommand{\p}[1]{(\ref{#1})}
\begin{document}
\renewcommand{\thefootnote}{\fnsymbol{footnote}}
\begin{center}
{\large\bf
On a ZERO CURVATURE REPRESENTATION
for BOSONIC STRINGS and $P$-BRANES}

\renewcommand{\thefootnote}{\dagger} \vspace{0.2cm}

{\bf Igor A. Bandos,}

\vspace{0.2cm}
{\it NNC Kharkov Institute of Physics and Technology} \\
{\it 310108, Kharkov, Ukraine}\\
e-mail:  kfti@rocket.kharkov.ua

\vspace{0.2cm}
\end{center}

{\small
\begin{quotation}
It is shown that a zero curvature
representation for $D$-- dimensional $p$--brane equations of motion
originates naturally in the geometric (Lund-- Regge-- Omnes) approach.

To study the possibility to use this zero curvature representation for
investigation of nonlinear equations of $p$-- branes, the simplest case
of $D$-- dimensional string ($p=1$) is considered.

The connection is found between the $SO(1,1)$ gauge (world--sheet
Lorentz) invariance of the string theory with a
nontrivial dependence on a spectral parameter of the
Lax matrices associated with the nonlinear equations
describing the embedding of a string world sheet into flat $D$--
dimensional space -- time. Namely, the spectral parameter can be
identified with a parameter of constant $SO(1,1)$ gauge transformations,
after the deformation of the Lax matrices has been performed.

\end{quotation} }

PACS: 10, 11.15-q, 11.17+y
\setcounter{page}1
\renewcommand{\thefootnote}{\arabic{footnote}}
\setcounter{footnote}0

\newpage

\medskip
{\large \bf Introduction}

In the recent paper by J. Hoppe
\cite{hoppe95} a zero curvature representation   was constructed for the
nonlinear equations of motion of bosonic $(D-2)$--branes moving in
$D$--dimensional space-- time.  Earlier a Lax pair
representation for a particular case of $D=4$ membrane was built in Ref.
\cite{hoppe}.  This zero curvature representation and the associated
linear system (without spectral parameter)
were used in Ref. \cite{hoppe95} for constructing the
nonlocal conserved charges.

To emphasize the importance of this result, let us note, that
the finding of a zero curvature representation for a self--dual
Yang--Mills equation inspires the discovery of the correspondence of
self--dual gauge fields with complex holomorphic vector bundles
\cite{ward} and, hence, was among the achievements which had led to the
significant progress in instanton and monopole physics as well as to the
development of related fields in mathematics (see, for example,
\cite{atiyah} and refs therein).
The problem of constructing a zero curvature
representation for a given system of (exactly solvable)
nonlinear equations was considered previously
from the mathematical point of view (see, for example \cite{marvan}).

In order to write the
$(D-2)$--brane nonlinear equations in the form of zero curvature
representation, the gauge was fixed  with respect to world surface
diffeomorphism as well as with respect to Lorentz transformations and the
pure spatial form of the extrinsic geometry formalism was developed in
\cite{hoppe95}.  However, as it was mentioned already in \cite{hoppe95},
the lack of manifest Lorentz invariance prohibits the use of the
zero curvature representation
in full measure.

\medskip

In the present paper it is shown that the
relativistic
invariant counterpart of the Hoppe's construction  \cite{hoppe95}
appears in a natural way in the
geometric approach \cite{lr} -- \cite{bpstv} for bosonic $p$--
branes\footnote{ $p$ is the number of space-- like dimensions of the world
volume of the extended object ($p=1$ for string, $p=2$ for membrane
etc.)}.
This zero curvature representation
\begin{equation}\label{MC0}
d\Omega ^{\underline a\underline b} -
\Omega ^{\underline a}_{~\underline c} \Omega ^{\underline c \underline b}
= 0,
\end{equation}
$$
\underline{a} , \underline{b}, \underline{c} = 0, 1, ..., (D-1)
$$
is formulated in terms of
$so(1,D-1)$ valued  $1$-- forms (connections)
\footnote{Here and below $d$ is the right external differential and the
product of forms is supposed to be external one, i.e.
$\Om_r \Om_q = (-1)^{qr} \Om_q\Om_r~,~~$
$~d(\Om_r\Om_q)=
\Om_rd\Om_q+ (-1)^q d\Om_r\Om_q$
for a product of any r-- and q-- forms.}
\begin{equation}\label{Cfv0}
\Om^{\underline{a}\underline{b}} = -
\Om^{\underline{b}\underline{a}} =
\left(
\matrix{ \Om^{ab} & \Om^{aj} \cr
       - \Om^{bi} & \Om^{ij} \cr}
        \right)
\end{equation}
$$
a, b, c = 0,1,..., p ~~~~~~~~~~ i, j = 1,..., (D-p-1)
$$
expressed through an intrinsic world volume vielbein matrix
$e^{~a}_{m}$
($e^a = d\xi^m e_m^{~a}, ~~~m=0,1,...,p$), symmetric and traceless second
fundamental form matrix
$({\cal K}^i)^{ba}~~$
($({\cal K}^i)^{ba} = ({\cal K}^i)^{ab}$,
$({\cal K}^i)^{ba} \eta_{ba} = 0$)
and world volume gauge fields $\Om^{ij}_m$
\begin{equation}\label{eome}
\Omega^{bc}
= e^a (e^{m}_{a} e^{nb} \partial_{{[m}} e_{{n]}}^c
- e^{mb} e^{nc} \partial_{{[m}} e_{{n] a}}
+ e^{mc} e^{{n}}_{{a}} \partial_{{[m}} e_{n]}^b )
\end{equation}
\begin{equation}\label{2ffi0}
\Om^{ai} = e_{b} {\cal K}^{ba~i}
\end{equation}
\begin{eqnarray}\label{gaugef}
\Om^{ij} = d\xi^m \Om^{ij}_m
\end{eqnarray}
It has manifest
global $SO(1,D-1)$ (Lorentz) and local (gauge) $SO(1,p) \otimes
SO(D-p-1)$  symmetries and it is suitable for a
description of any bosonic $p$--brane moving in
 space time of any dimension $D >p$.

It can be shown that the zero curvature representation \p{MC0} --
\p{gaugef} is equivalent to the set of the Peterson-- Codazzi, Gauss and
Rici equations of the classical theory of the surfaces \cite{Ei,barnes}.
In this sense, it is not new.

Our proposition is
to use the manifest $SO(1,p) \otimes SO(D-p-1)$ gauge symmetry of the
considered zero-- curvature representation \p{MC0} -- \p{gaugef} for
investigation of the nonlinear equations of motion for $p$-- branes with
$p\geq 2$ \footnote{The standard way to deal with the geometric approach
equations consists in attempts to reduce the number of functions involved
as well as the number of equations \cite{barnes}.
Thus, in Ref. \cite{zhelt} the number of gauges were fixed and the
geometric approach equations for $D$-- dimensional bosonic string were
reduced to a system of nonlinear equations for $(D-2)$ physical
(transverse) degrees of freedom.

Let us note here that geometric approach
had been studied before for string theory mostly. }.

The first problem we would like to investigate is whether it is
possible to introduce a nontrivial dependence on a spectral parameter
\cite{int,faddeev} into the connections \p{Cfv0} - \p{gaugef}
using the manifest $SO(1,p) \otimes SO(D-p-1)$ gauge invariance
of the zero curvature representation \p{MC0}-- \p{gaugef}
for $p$-- brane equations.

It is reasonable
to begin the investigation of the above problem with studying
the simplest case of string ($p=1$) in diverse dimensions $D \geq 3$.
This is the main subject of the present paper.

The equations of motion for string theory become linear if
a suitable gauge is fixed. However, the embedding of string world-- sheet
is described by a system of
nonlinear equations
\cite{lr,barnes,zhelt}
in the frame of geometric approach.
For $D=3$ string this is the exactly solvable Liouville equation
\cite{das,kulish}.

The components
$
\Om_m^{\underline{a}\underline{b}} =
(\Om_\tau ^{\underline{a}\underline{b}},
\Om_\sigma^{\underline{a}\underline{b}}) \equiv
(1/2(\Om_{(++)}^{\underline{a}\underline{b}} +
\Om_{(--)}^{\underline{a}\underline{b}}),
1/2(\Om_{(++)}^{\underline{a}\underline{b}} -
\Om_{(--)}^{\underline{a}\underline{b}}))$
of the connection $1$-- form \p{Cfv0}
$\Om^{\underline{a}\underline{b}} =
d\xi^m \Om_m^{\underline{a}\underline{b}}
= d\tau \Om_\tau^{\underline{a}\underline{b}} +
d\sigma \Om_\sigma^{\underline{a}\underline{b}} =
d\xi^{(\pm\pm)}\Om_{(\pm\pm)}^{\underline{a}\underline{b}}$
are the counterparts of $U$ and $V$ matrices \cite{int,faddeev} or,
equivalently, of $L$ and $A$ operators (see \cite{das,kulish} and Refs.
therein), which are known for all exactly solvable equations.
They can be used to define the associated linear system
\begin{equation}\label{aslin}
\hat{D} \Psi^{\underline{a}}
\equiv d\Psi^{\underline{a}} - \Psi^{\underline{b}}
\Om^{~\underline{a}}_{\underline{b}} = 0 ,
\end{equation}
However, for all known exactly solvable systems $L$ and $A$
operators do depend on a spectral parameter in a nontrivial way. This
dependence is
necessary to use the full power of the Inverse Scattering Method (see
\cite{int,das,faddeev} and Refs. therein) and it can be used to obtain
infinite number of conserved currents \cite{int,hoppe95}.

The considered zero curvature representation \p{MC0}-- \p{gaugef} involves
the connections \p{Cfv0}-- \p{gaugef} being independent on spectral
parameter.
However, it has the manifest
$SO(1,p) \otimes SO(D-p-1)$ gauge invariance instead
($SO(1,1) \otimes SO(D-2)$ for the string case $p=1$).

Here we will show that the nontrivial dependence of $U = \Om_\s$ and
$V=\Om_\tau$ matrices on the spectral parameter $\a$ can be derived
by the use of the $SO(1,1)$ gauge symmetry of the zero curvature
representation.

For the case $D=3$ our prescription reproduces the
standard associated linear system with spectral parameter for nonlinear
Liouville equation \cite{kulish} as well as one derived from the
continuous limit of the
massless limit of the lattice version of the Sine-- Gordon model
in the recent paper \cite{faddeev}.

The prescription makes possible get the similar associated linear
systems for nonlinear equations describing
the embedding of the string world sheet into space-- time of high
dimensions $D > 4$, as well as for $n=(1,0)$ and $n=(1,1)$ supersymmetric
Liouville systems \cite{bsv1} describing $N=1$ and $N=2$ superstrings in
$D=3$.

\medskip

\section{Maurer--Cartan
equations as a zero curvature representation for
$p$-brane equations of motion}

 In the standard $p$--brane formulation \cite{dirac} the
(minimal) embedding of the world volume into a flat $D$-- dimensional
target space is described by the coordinate functions
\begin{equation}\label{X(xi)}
X^{\underline m} = X^{\underline m}(\xi^m) ,
\qquad
\underline{m} = 0, 1, \ldots, (D-1), \qquad m = 0, 1, \ldots p, \qquad
\end{equation}
satisfying the equation
\begin{equation}\label{eqm}
\partial_m ( \sqrt{|g|} g^{mn} \partial_n X^{\underline m}) = 0 ,
\qquad
g_{mn} =
\partial_m X^{\underline m} \partial_n X^{\underline n}
\eta_{\underline{m}\underline{n}} ,
\end{equation}
which becomes nonlinear for $p>1$.
Here $g_{mn}$ is the induced world volume metric
and $\eta_{\underline{m}\underline{n}}$ is the flat target space-- time
metric
$
\eta_{\underline{m}\underline{n}} = diag (+,-,\ldots, -)
$.

\medskip

In the geometric approach \cite{lr,barnes,zhelt}
$p$-- brane theory  is described by the world-- volume vielbein form
$e^a = d\xi^m e_m^{~a}$
and the $so(1,D-1)$ valued Cartan $1$-- form \p{Cfv0}
\begin{equation}\label{Cfv}
\Om^{\underline{a}\underline{b}} = -
\Om^{\underline{b}\underline{a}} =
\left(
\matrix{ \Om^{ab} & \Om^{aj} \cr
       - \Om^{bi} & \Om^{ij} \cr}
        \right)
\end{equation}
which satisfy the
Maurer-- Cartan equation \p{MC0}
\begin{equation}\label{MC}
d\Omega ^{\underline a\underline b} -
\Omega ^{\underline a}_{~\underline c} \Omega ^{\underline c \underline b}
= 0,
\end{equation}

After the natural $SO(1,p) \otimes SO(D-p-1)$ invariant splitting
\p{Cfv} we can obtain
the forms $\Om^{ab},~ \Om^{ij}$ having the properties of
$SO(1,p)$ and $SO(D-p-1)$  connections as well as the covariant form
$\Om^{ai}$ being the vielbein of the coset
$SO(1,D-1) / (SO(1,p) \otimes SO(D-p-1))$.
Maurer-- Cartan equation \p{MC} splits into the following equations
for the forms $\Om^{ab},~ \Om^{ij}$ and $\Om^{ai}$
\begin{equation}\label{PCg}
{\cal D}\Omega ^{ai} \equiv
d\Omega ^{ai} -
\Omega ^{a}_{~b} \Omega ^{bi} +
\Omega ^{aj} \Omega ^{ji} = 0,
\end{equation}
\begin{equation}\label{Gg}
{R}^{ab}(d,d)=
d\Omega ^{ab} -
\Omega ^{a}_{~c} \Omega ^{cb} =
\Omega ^{ai} \Omega^{bi},
\end{equation}
\begin{equation}\label{Rg}
{R}^{ij}(d,d)=
d\Omega ^{ij} +
\Omega ^{ij^\prime} \Omega^{j^\prime j} = -
\Omega ^{ai} \Omega^{~j}_{a},
\end{equation}
which give rise to the Peterson--Codazzi, Gauss and Ricci
equations of the surface embedding theory, respectively
 \cite{barnes,bpstv}.

The $1$-- form variables  $e^a$ and $\Om^{\underline{a} \underline{b} }
= d\xi^m \Om^{\underline{a} \underline{b}}_m $
are related by the equations
\begin{equation}\label{eomi}
e_a \Om^{ai} = 0 ,
\end{equation}
\begin{equation}\label{mini}
e^{~m}_{a} \Om^{~ai}_{m} = 0 ,
\end{equation}
\begin{equation}\label{tor}
T^a \equiv {\cal D} e^{a} \equiv
de^a - e_b \Om^{ba} = 0 ,
\end{equation}
The left hand parts of
Eqs. \p{eomi}, \p{tor} are differential $2$ forms,
while Eq.\p{mini} is written in terms of components $\Om^{ai}_m$ of
the form $\Om^{ai} = d\xi^{m} \Om^{ai}_m$ and inverse vielbein matrix
$e^m_a$.

Eq. \p{tor} means that the world volume vielbein
is covariantly constant with respect to the parallel transport
defined by the connection $\Om^{ab}$.

Eq. \p{mini}
is the only equation of the set \p{MC} -- \p{tor} which
has dynamical content. From the standpoint of surface theory, it
defines the embedding as being minimal one.
On the other hand it can be reduced to Eq. \p{eqm} if eliminating  the
auxiliary variables.

In order to show this\footnote{
The detailed consideration of
relation between the standard $p$--brane description \p{eqm} and that of
the geometric approach is given in Ref. \cite{bpstv}, where the
supersymmetric generalization of the geometric approach is performed.}
one needs,  first of all, to  solve the Maurer-- Cartan equations \p{MC}
($=$ \p{PCg} - \p{Rg}) expressing $so(1,D-1)$ valued connection
$\Om^{\underline{a}\underline{b}}$ in terms of
$SO(1,D-1)$ valued matrix
\begin{equation}\label{split}
 ||u_{\underline{m}} ^{\underline{a}} ||
 \equiv || ( u_{\underline{m}}^{~a}, u_{\underline{m}}^i)||
 ~~~ \in ~~~  SO(1,D-1)
 \end{equation}

\begin{equation}\label{orthon}
\Leftrightarrow
u_{\underline{m}}^{\underline{a}}
 u^{\underline{m}\underline{b}} =
 \eta^{\underline{a}\underline{b}} = diag (+,-,...,-)
 \Leftrightarrow
\cases{
u_{\underline{m}}^{a} u^{\underline{m}b} = \eta^{ab} , \cr
 u_{\underline{m}}^{a} u^{\underline{m}j} = 0 ,  \cr
 u_{\underline{m}}^{i} u^{\underline{m}~j} = - \d^{ij} \cr }
\end{equation}
The solution has the form
\begin{equation}\label{Cf}
\Om^{\underline{a}\underline{b}} = u^{~\underline{a}}_{\underline{m}}
d u^{\underline{b} \underline{m}}
 \Leftrightarrow
\cases{
 \Om^{ai} = u^{a}_{\underline m} d u^{\underline m~i} , \cr
\Om^{ab} = u^{a}_{\underline m} d u^{\underline m~b} , \cr
\Om^{ij} = u^{i}_{\underline m} d u^{\underline m~j} \cr }
\end{equation}
The vectors $u^{a}_{\underline m}$ and $u^{i}_{\underline m}$
can be considered as $(p+1) \times  D$ and $(D-p-1) \times D$
rectangular blocks of the Lorentz group valued matrix
\p{split}. They can be identified with
${{SO(1,D-1)} \over {SO(1,p) \otimes SO(D-p-1)}}$
Lorentz harmonics or, equivalently, with multidimensional
generalization of Newman-- Penrose dyades (see \cite{bz,bpstv} and refs.
therein).

Using the unity matrix decomposition in terms of moving frame vectors $u$
\begin{equation}\label{unity}
\d_{\underline{m}}^{\underline{~n}} =
 u_{\underline{m}}^{\underline{a}} u^{\underline{n}}_{\underline{a}}
 = u_{\underline{m}}^{a} u^{\underline{n}}_a -
 u_{\underline{m}}^{i} u^{\underline{n}~i}
\end{equation}
and Eqs. \p{Cf}, Eqs.
\p{tor} and \p{eomi} may be combined to form one equation
\begin{equation}\label{deu}
d(e^a u^{\underline{m}}_{a}) = 0
\end{equation}
The closed forms
$e^a u^{\underline{m}}_{a}$  \p{deu}
can be represented locally as external
differentials of some scalar (with respect to world volume) functions
\begin{equation}\label{eudX}
e^a u^{\underline{m}}_a = dX^{\underline{m}}
\end{equation}
In such a way the coordinate functions  \p{X(xi)}
appear in geometric approach \cite{lr,barnes}.

Taking into account
Eq. \p{Cf} and the relation
$$
u^{\underline{m}a} = e^{an} \partial_n X^{\underline{m}},
$$
which is equivalent to \p{eudX},
Eq. \p{mini} acquires the form of Eq.\p{eqm} with
$g^{mn} = e^{m}_a \eta^{ab} e^{n}_b $.

\bigskip

To get {\sl a zero curvature representation}
for arbitrary bosonic
$p$--brane on the base of the Maurer-- Cartan equation \p{MC},
we shall only prove the following {\large \it statement}:
\\ {\sl All the equations of
the geometric approach besides those included in the Maurer-- Cartan
equation \p{MC} can be solved with respect to the
Cartan forms $\Om^{ab}$ and $\Om^{ai}$.

\medskip

The {\large \it proof} of this statement is very simple:

Eq. \p{tor} can be solved with respect to
$\Om^{ab}= d\xi^m \Om^{ab}_m $.
Such solution is
known from General Relativity and has the form
\p{eome}.

General solution of Eqs. \p{eomi}, \p{mini}
has the form \p{2ffi0}
\begin{equation}\label{2ffi}
\Om^{ai} = e_{b} {\cal K}^{ba~i}
\end{equation}
where symmetric and traceless matrices
$({\cal K}^i)^{ba}~~$
($({\cal K}^i)^{ba} = ({\cal K}^i)^{ab}$,
$({\cal K}^i)^{ba} \eta_{ba} = 0$)
are  related to the second fundamental form
(see, for example,  \cite{bpstv}).

\medskip

Thus, the pull--backs of the Cartan forms
$\Om^{ai}$ and $\Om^{ab}$
are expressed in terms of second fundamental form
matrix ${\cal K}^{ab~i}$ and intrinsic world volume vielbein matrix
$e^{~a}_{m}$ by  Eqs. \p{eome}, \p{2ffi}.
The pull-- back of $SO(D-p-1)$ connection form
$\Om^{ij} = d\xi^m \Om_m^{ij}$ remains independent and gives rise
to the world volume gauge fields  $\Om_m^{ij} = \Om_m^{ij} (\xi^n)$
\footnote{see \cite{zhelt} for reformulation of string theory in terms of
two dimensional gauge fields.}.

\medskip

The very significant fact is that
{\sl all the equations for these variables are included
into the set of Maurer-- Cartan  equations \p{MC} (or \p{PCg} --\p{Rg})
having the form of zero curvature condition.}

\medskip

It is remarkable that the  zero curvature representation under
consideration can be written in spinor notations
\begin{equation}\label{MCs}
d \Om^{ ~\underline{\b }} _{\underline{\a }}
- \Om^{~\underline{\g }} _{\underline{\a }}
\Om^{~\underline{\b }}_{ \underline{\g }} = 0 ,
\end{equation}
where
$\Om^{ ~\underline{\b }} _{\underline{\a }} $ is
$spin(1,D-1)$ valued Cartan form. It is
 related to $so(1,D-1)$ valued $1$-- form
$\Om^{ \underline{b }\underline{a }} $
by $D$-- dimensional $\Gamma$-- matrices
\begin{equation}\label{Cfsg}
\Om^{ ~\underline{\b }}_{\underline{\a }}
\propto
\Om^{ \underline{b }\underline{a }}
(\Gamma_{ \underline{b }\underline{a }})
^{ ~\underline{\b }}_{\underline{\a }}
\propto
\Om^{ \underline{b }\underline{a }}
(\Gamma_{ \underline{b }} \Gamma_{\underline{a }})
^{ ~\underline{\b }}_{\underline{\a }}
\end{equation}
The  coefficients in Eq. \p{Cfsg} are dependent on the number of target
space-- time dimensions $D$.

The associated linear system,  which
reproduce the considered zero curvature representation
 \p{MC}, \p{Cfv}, \p{eome}, \p{2ffi} as integrability conditions, has the
 evident form  \p{aslin}
 \begin{equation}\label{aslin1}
 \hat{D} \Psi^{\underline{a}}
\equiv d\Psi^{\underline{a}} - \Psi^{\underline{b}}
\Om^{~\underline{a}}_{\underline{b}} = 0 ,
\end{equation}
with
$\Om^{~\underline{b}} _{\underline{a}} =
\eta_{\underline{a}\underline{c}} \Om^{\underline{c}\underline{a}} $
determined by \p{Cfv}, \p{eome}, \p{2ffi}.
Otherwise, it can be identified with equivalent form
$d u^{\underline{a} \underline{m}} =
 u^{~\underline{b} \underline{m}}
\Om^{~\underline{a}}_{\underline{b}}$
of Eq. \p{Cf} or with spinor counterparts of the above equations.

\medskip

In conclusion, the relativistic invariant zero curvature representation
for $D$- dimensional $p$-- brane equations of motion is given by the
Maurer-- Cartan equation \p{MC} for $so(1,D-1)$ valued Cartan $1$-- forms
\p{Cfv} with
$\Om^{ai} = d\xi^m \Om^{ai}_{m} (\xi )$ and
$\Om^{ab} = d\xi^m \Om^{ab}_{m} (\xi )$  specified by Eqs. \p{2ffi}
and \p{eome}.

\medskip

Of course, this zero curvature representation is equivalent to the set of
Peterson-- Codazzi, Gauss and Rici equations and, in this sense,
is not new. Our proposition is to use its manifest global $SO(1,D-1)$ and
local $SO(1,p) \otimes SO(D-p-1)$ symmetries for investigation of the
$p$-- brane equations of motion.

\medskip

One of the  intriguing possibilities is to use these manifest
symmetries
to include a spectral parameter into
the associated linear system
\p{aslin}.

If this is impossible for a general case, then the problem
can be reformulated as
one of classification of all the particular cases of
$p$-- brane motions in $D$-- dimensional space time, where the
introduction of spectral parameter is possible.

\medskip

Below we will consider the zero curvature representation
\p{MC} for nonlinear
equations describing bosonic string theory in geometric approach.
We will present an explicit prescription for inclusion of spectral
parameter into the associated linear system \p{aslin1}
starting from $SO(1,1)$ gauge symmetry of \p{MC}.
For the simplest case of string in $D=3$ dimensional space-- time,
which is described by nonlinear Liouville equation, we
reproduces in such a way the standard associated linear system with
spectral parameter \cite{kulish} as well as one derived from the lattice
limit in Ref. \cite{faddeev}.

\medskip

\section{ Geometric approach for bosonic string.}

\begin{center}
{\bf \Large $SO(1,1)$ gauge invariance and spectral parameter}
\end{center}

For $p=1$ case
the world sheet vielbein  can be
splitted into $SO(1,1)$ covariant light-- like components
$ e^a = (e^{++}, e^{--})$.

The set of $SO(1,D-1)$ Cartan forms
$\Om^{\underline a \underline b} \equiv
- \Om^{\underline b \underline a}$
\begin{equation}\label{3cfs}
\Om_{\underline a}^{~\underline b} =
\eta_{\underline a \underline c} \Om^{\underline c \underline b}
\equiv \left(
\matrix{\Om^{(0)} & 0 &  2^{-1/2} \Om^{--i} \cr
        0 & - \Om^{(0)} & 2^{-1/2} \Om^{++i} \cr
        2^{-1/2} \Om^{++j} & 2^{-1/2} \Om^{--j} & \Om^{ij} \cr}
        \right)
\end{equation}
splits naturally into the $SO(1,1)$ connection $\Om^{ab} \propto
\epsilon^{ab} \Om^{(0)}$, $SO(D-2)$ connection $\Om^{ij}$
and two ${{SO(1,D-1)} \over {SO(1,1) \times SO(D-2)}}$ vielbein forms
$\Om^{ai} = (\Om^{++i}, \Om^{--i})$.

The parts \p{PCg}, \p{Gg} and \p{Rg}
of the Maurer--Cartan equation \p{MC}
have the form
\begin{equation}\label{3pc}
{\cal D}\Om^{\pm\pm i}
\equiv d\Om^{\pm\pm i} \pm \Om^{(0)} \Om^{\pm\pm i}
+ \Om^{\pm\pm j} \Om^{ji} = 0
\end{equation}
\begin{equation}\label{3g}
{\cal F} \equiv
d\Om^{(0)}
= {1 \over 2} \Om^{--i} \Om^{++i} ,
\end{equation}
  \begin{equation}\label{Rst}
{R}^{ij}(d,d)=
d\Omega ^{ij} +
\Omega ^{ij^\prime} \Omega^{j^\prime j}
= -  \Om^{--[i} \Om^{++j]} ,
\end{equation}
and
Eqs. \p{tor}, \p{eomi} and \p{mini}
take the forms
\begin{equation}\label{3tor}
T^{\pm\pm} \equiv {\cal D} e^{\pm\pm} = de^{\pm\pm} \pm \Om^{(0)}
e^{\pm\pm} = 0 ,
\end{equation}
\begin{equation}\label{3sym}
e^{++} \Om^{--i} + e^{--} \Om^{++i} = 0 ,
\end{equation}
\begin{equation}\label{3min}
e^{m}_{++} \Om^{++i}_{m} + e^{m}_{--} \Om^{--i}_{m} = 0
\end{equation}

\subsection{ String in $D=3$ and Liouville equation.}

For the simplest case of $D=3$ bosonic string internal index $i$ has
only one value $i = \perp $ and internal connection $\Om^{ij}$ is absent.

Eqs. \p{3sym} and \p{3min} have  simple solution
\begin{equation}\label{om++}
\Om^{++} = e^{--} \Om^{~++}_{--} ,
\qquad \Om^{--} = e^{++} \Om^{~--}_{++}
\end{equation}
$ \Om^{~++}_{--}$ and  $\Om^{~--}_{++}$ can be regarded as
the nonvanishing components of symmetric and traceless second fundamental
form matrix ${\cal K}^{ab}$ \p{2ffi}.

Then we can solve Eq. \p{3pc} with respect to
(induced) spin connection $\Om^{(0)}$
\begin{equation}\label{om0}
\Om^{(0)} \equiv e^{\pm\pm} \Om^{(0)}_{\pm\pm} =
{1\over 2} e^{++} (\Om^{~++}_{--})^{-1}
e^{m}_{++} \partial_m \Om^{~++}_{--} -
{1\over 2} e^{--} (\Om^{~--}_{++})^{-1}
e^{m}_{--} \partial_m \Om^{~--}_{++}
\end{equation}

Using \p{om0}, Eq. \p{3tor} can be written in the
form
$$
 d((\Om^{~++}_{--})^{1/2} e^{--}) = 0 , \qquad
d((\Om^{~--}_{++})^{1/2} e^{++}) = 0  ~~~$$
or
$$ e^{--} = (\Om^{~++}_{--})^{-1/2} d\xi^{(--)} , \qquad
 e^{++} = (\Om^{~--}_{++})^{-1/2} d\xi^{(++)} ~~~$$
This means the conformal flatness of any  $2$--dimensional geometry.
Further we will consider
$\xi^{++}, ~\xi^{--}$ as functions of world sheet
coordinates $\xi^{\mu}=(\tau , \sigma)$. By this we exclude
from the consideration the possible
contributions from a nontrivial world sheet topology.
The influence of such inputs is an interesting
problem for further investigations.

It is convenient to use $\xi^{(\pm\pm)}$ as the local world sheet
coordinates (and thus to fix the gauge with respect to
reparametrization symmetry) and rewrite the expressions for the Cartan
form pull backs in term of the holonomic basis
$d\xi^{(\pm\pm)}$
of the cotangent space
($\partial_{(\pm\pm )} = {\partial
\over {\partial \xi^{(\pm\pm)}}}$)
\begin{equation}\label{om++h}
\Om^{++} =
exp(W - L) d\xi^{(--)}
\end{equation}
\begin{equation}\label{om--h}
\Om^{--} = exp(W + L) d\xi^{(++)}
\end{equation}
\begin{equation}\label{om0h}
\Om^{(0)} = (d\xi^{(++)} \partial_{(++)} -
d\xi^{(--)} \partial_{(--)}) W - dL
\end{equation}
Here the gauge
invariant degree of freedom of the fields $\Om^{~++}_{--}$ and
$\Om^{~--}_{++}$ is denoted by $W$ $$~exp(4W) \equiv
\Om^{~++}_{--} \Om^{~--}_{++} ,~$$  and another (pure
gauge) degree of freedom is denoted by $L$ $$~exp(4L)
\equiv  \Om^{~--}_{++} /  \Om^{~++}_{--}$$

The Gauss equation \p{3g}
for the forms \p{om++h} -- \p{om0h} gives rise to the
Liouville equation
\begin{equation}\label{Leq}
\partial_{(++)}\partial_{(--)} W = {1 \over 4} exp(2W)
\end{equation}
while the Peterson-- Codazzi equations \p{3pc} are satisfied identically.

The field $L$  remains unrestricted. This reflects the local
$SO(1,1)$ (world sheet Lorentz) invariance of the string theory.

Thus we have reproduced the known result \cite{lr,barnes}:
in the framework of geometric approach $D=3$ string is described by
Liouville equation \p{Leq}.

However, in the proposed approach Liouville equation appears in the form
of manifestly $SO(1,1)$ gauge invariant zero curvature representation
\p{MC} (or \p{MCs}) for the $so(1,2)$ valued Cartan forms
\p{Cfv} (or \p{Cfsg}), \p{om++h} -- \p{om0h}.

Below we will
find an explicit relation between the spectral
parameter of the linear system associated with the Liouville equation
\cite{kulish} and $SO(1,1)$ gauge invariance of the zero curvature
representation ( i.e.  world sheet Lorentz invariance
of string theory). Namely, we will show that the spectral parameter can be
identified with a parameter of constant $SO(1,1)$ transformations after
a deformation of the Lax matrices (i.e. $SO(1,2)$  connections) has been
performed.

\medskip

\subsection{Spectral parameter and gauge invariance.}

Below it is convenient
to use a spinor representation \p{MCs}
of the Maurer -- Cartan equation \p{MC}.
For $D=3,$ case $so(1,D-1) =so(1,2) = sl(2,R)$ and
$sl(2,R)$ valued connection
$\Om_{\underline \a}^{~\underline \b}$
has the form
\begin{equation}\label{Oms}
\Om_{\underline \a}^{~\underline \b} =  {1 \over 2}
\left(
      \matrix{ \Om^{(0)} & \Om^{--}  \cr
               \Om^{++} & - \Om^{(0)} \cr}
                                          \right)
\end{equation}
where $\Om^{(0)},~ \Om^{\pm\pm }$ are still defined by eqs.
\p{om0h}, \p{om++h}, \p{om--h}.

The Liouville equation \p{Leq} appears as a diagonal element of the matrix
equation \p{MCs}.

\medskip

In order
to reproduce the associated linear system
(i.e. $U$ and $V$ operators \cite{int,faddeev}) being dependent
nontrivially on a spectral parameter \cite{kulish,faddeev}
the following trick shall be done.

Let us  {\it deform} $SL(2,R)~ (= SO(1,2))$ connection
\p{Oms}  replacing the coset
form $\Om^{++}$ by its sum with the other coset form
$\Om^{--}$ multiplied by a function  $f^{~++}_{--}(\xi)$:  $~~~ \Om^{++}
\rightarrow (\Om^{++} + f^{~++}_{--} \Om^{--}) $

\begin{equation}\label{deform}
\Om_{\underline \a}^{~\underline \b}  ~~~\rightarrow ~~~
\Om_{\underline \a}^{~\underline \b} (f) \equiv
{1 \over 2}
\left(
      \matrix{ \Om^{(0)} & \Om^{--}  \cr
               \Om^{++} + f^{~++}_{--} \Om^{--} & - \Om^{(0)} \cr}
                                          \right)
\end{equation}
In general case the  curvature of the deformed connections
\p{deform} does not vanish
\begin{equation}\label{Mcdef}
d\Om_{\underline \a}^{~\underline \b} (f) -
\Om_{\underline \a}^{~\underline \g} (f)
\Om_{\underline \g}^{~\underline \b} (f) =
{1 \over 2}
\Om^{--} {\cal D} f^{~++}_{--}
\left(
      \matrix{ 0 & 0 \cr
               1 & 0 \cr}
                                            \right)
\end{equation}
In Eq. \p{Mcdef} the Liouville equation \p{Leq} as well as the explicit
form of $\Om^{\pm\pm}, \Om^{(0)}$ \p{om++h} -- \p{om0h} are taken into
account.

The class of the deformed connections with zero curvature is defined by
the functions $f^{~++}_{--}(\xi)$ which satisfy
\footnote{Eq. \p{df} means that the deformed connection $\Om (f)$ is
related with $\Om = \Om (0)$ by a local $SL(2,R)$ transformation.}
\begin{equation}\label{df}
\Om^{--} {\cal D} f^{~++}_{--}  \equiv
\Om^{--} (d f^{~++}_{--} - 2  f^{~++}_{--} \Om^{(0)}) = 0 ~~~
\Rightarrow ~~~ \partial_{(--)} ( e^{2(W+L)} f^{~++}_{--} ) = 0
\end{equation}
The general solution of Eq. \p{df} has the form
\begin{equation}\label{fh}
f^{~++}_{--} (\xi^m) = h(\xi^{(++)}) e^{-2(W+L)}
\end{equation}
where $h$ is a function of $\xi^{(++)}$ only ($\partial_{(--)} h = 0$).
The later can be gauged to a constant using conformal symmetry of the
considered system. Thus we could consider \p{fh} with a constant $h$ as a
general solution of Eq. \p{df}.

\medskip

The simplest kind of deformations is characterized by a constant function
$f^{~++}_{--}$ ($~df^{~++}_{--}=0$). In this case Eq. \p{df} acquire the
form
       \begin{equation}\label{dfconst}
f^{~++}_{--} \Om^{(0)} \Om^{--} = f^{~++}_{--} d \Om^{--} = 0
       \end{equation}
If one fixes the gauge with respect to
 $SO(1,1)$ symmetry
\footnote{Note, that all elements of the
matrix \p{Oms} have well defined properties with respect to $SO(1,1)$
gauge transformations, which suggested to be conserved after
the deformation \p{deform}.}
\begin{equation}\label{gauge}
L(\xi) = - W(\xi) + ln \a ~~ ,
\end{equation}
where $\a$ is a constant, then the form
$\Om^{--}$ becomes closed
$$ \Om^{--} = \a
d\xi^{(++)}  ~~ \Rightarrow ~~ d \Om^{--} = 0 $$
Hence, the right hand side of eq. \p{Mcdef} vanishes for constant $f$.

In this gauge the flat $SO(1,2)$ connection \p{Oms} acquires the form
\begin{equation}\label{Omf}
\Om_{\underline \a}^{~\underline \b} (\a , f) =
\left(
   \matrix{
d\xi^{(++)} \partial_{(++)} W & { \a \over 2} d\xi^{(++)} \cr
{1 \over {2\a }} d\xi^{(--)} exp\{ 2W\} + f^{~--}_{++} { \a \over 2}
d\xi^{(++)} & - d\xi^{(++)} \partial_{(++)} W \cr} \right) \end{equation}
and the associated linear system with the spectral parameter $\a$
\begin{equation}\label{lin}
d\Psi_{\underline \a}
- \Om_{\underline \a}^{~\underline \b} (\a , f)
d\Psi_{\underline \b} = 0 ,
\end{equation}
{\bf coincides with one presented in} \cite{kulish}
for the choice of the constant
$
f^{~++}_{--} = - 1$
\footnote{
To get the associated linear system in the form presented in Ref.
\cite{kulish} the following redefinition of the variables and the
spectral parameter shall be done:
$\xi^{(--)} = \tau $, $\xi^{(++)} = 4 \s $, $W(\xi ) = \phi (\tau, \s )$,
$ - 2^{-1/2} \a = \l $.}.
The constants $f$ and $h$ \p{fh} are
related by $f = {h \over {\a ^2}}$.

If one comes back to the general solution \p{fh} of Eq. \p{df} and fixes
the gauge
\begin{equation}\label{gauge2}
L(\xi ) = ln \a = const
\end{equation}
with respect to $SO(1,1)$ symmetry, then the flat $SL(2,R)$ connections
\p{deform} takes the form
$$
\Om_{\underline \a}^{~\underline \b} (\a, h) = d\xi^{\pm\pm}
\Om_{(\pm\pm)\underline \a}^{~~~~~~\underline \b} (\a , h)
$$
\begin{equation}\label{faddeev}
\Om_{(++)}(\a , h)
= {1 \over 2} \left(
   \matrix{
\partial_{(++)} W &  \a e^W  \cr
{h \over {\a}} e^{-W} & - \partial_{(++)} W \cr} \right)  ,
\qquad
\Om_{(--)} (\a , h)
= {1 \over 2} \left(
   \matrix{
- \partial_{(--)} W & 0 \cr
{1 \over {\a }}  e^W  & \partial_{(--)} W \cr} \right)
\end{equation}
For the choice $h= - \a^2~~$ ($f^{~++}_{--} = - e^{-2W}$), the matrices
$U = 2i (\Om_{(--)} - \Om_{(++)})$,
$V = - 2i (\Om_{(--)} + \Om_{(++)})$ coincide with the Lax pair for
Liouville equation obtained in Ref. \cite{faddeev}
\footnote{The relation between the variables and the spectral parameter
of the present work  with ones from Ref. \cite{faddeev} are
$\xi^{\pm\pm} = -4 (\s \pm \tau )$, $W(\xi ) = -i \Phi (\s, \tau )$,
$\a = -i e^{- \l }$.}.
The {\sl nontrivial} dependence of these $U$ and $V$ matrices on
the spectral parameter $\a$ were used in \cite{faddeev} to apply the
Quantum Inverse Scattering Method \cite{faddeev1} for studying the quantum
Liouville system.

For both cases the parameter $\a$ of the global
$SO(1,1)$ transformations, which appears as the reminder of the gauge
$SO(1,1)$ symmetry of the Gauss and Peterson-- Codazzi equations \p{Gg},
\p{PCg}, becomes the spectral parameter of the associated linear system
\p{lin}.

Thus there is the close relation of the spectral parameter with the gauge
symmetries of the zero curvature representations.

However, we should stress, that this relation
should not be taken literally.
So, if one puts $f=h=0$ and identifies the spectral parameter $\a$ with
constant $SO(1,1)$ gauge transformation parameter, then Eq. \p{faddeev}
reproduces the $U$ and $V$ operators being dependent on the
spectral parameter $\a $ in a trivial way \cite{faddeev2,faddeev}.

So, the possibility of constructing the deformations similar to \p{deform}
without breaking the Maurer--Cartan equation \p{MCs} is the crucial
factor for a nontrivial introduction of a spectral
parameter into the associated linear system.

\medskip

In the same manner the spectral parameter can be introduced into the
linear system associated with nonlinear system of equations describing
minimal embedding of string world sheet into the flat space time of any
dimension $D>3$, which we will describe shortly in the next section.

\subsection{ Bosonic string in diverse dimensions and
the sets of nonlinear equations}

As in the simplest $D=3$ case, all equations for the
 bosonic string in $D \geq 4$
can be represented as
the Maurer-- Cartan equation \p{MC}  for 1-- form \p{3cfs} with
\begin{equation}\label{Dom++h}
\Om^{++i} = d\xi^{(--)} exp(W + L) G^{ij}_L M^j
\end{equation}
\begin{equation}\label{Dom--h}
\Om^{--i} = d\xi^{(++)} exp(W - L) G^{ij}_R N^{j}
\end{equation}
\begin{equation}\label{Dom0h}
\Om^{(0)} = (d\xi^{(++)} \partial_{(++)} -
d\xi^{(--)} \partial_{(--)}) W - dL
\end{equation}
\begin{equation}\label{Domijh}
\Om^{ij} =
d\xi^{(--)} \partial_{(--)} G^{ik}_R  G^{jk}_R +
d\xi^{(++)} \partial_{(++)} G^{ik}_L  G^{jk}_L
\end{equation}
Here $M$ and $N$ are $SO(D-2)$ vector fields of opposite chirality
\begin{equation}\label{chiral}
\partial_{++} M^i  = 0 \qquad  \partial_{--} N^i = 0 ,
\end{equation}
whose norms can be fixed to be unity
\begin{equation}\label{unit}
M^i M^i = 1 , \qquad N^i N^i = 1 \qquad
\end{equation}
using the conformal symmetry of the system under consideration.
$G_R, ~G_L$ are orthogonal ($SO(D-2)$ valued) matrix fields
\begin{equation}\label{Gorthog}
G^{ik}_R  G^{jk}_R \equiv (G_R G_R^T)^{ij} = \d^{ij} \qquad
G^{ik}_L  G^{jk}_L \equiv (G_L G_L^T)^{ij}= \d^{ij}
\end{equation}

The field $W$ as well as the matrix valued field
$G^{ji} \equiv G_L^{ki} G_R^{kj}$ are invariant under the natural $SO(1,1)
\otimes SO(D-2)$ gauge symmetry of the considered system.

\medskip

The system of nonlinear equations for the above fields, which follows
from Eq. \p{MC}, is rather complicated. It involves, besides the chirality
conditions for $M$ and $N$ fields \p{chiral}, the Liouville-- like
equation
$$\partial_{(++)} \partial_{(--)} W =
{1 \over 4} e^{2W} N^i G^{ij} M^j
$$
 and the $\sigma$--model--
like equation for the orthogonal matrix fields
$$\partial_{(--)} (\partial_{++}G_L ~G_L^{T})^{ij} -
\partial_{++} (\partial_{--}G_R ~G_R^{T})^{ij} +
[\partial_{++} G_L G_L^{T} , \partial_{--}G_R G_R^{T}]^{ij}
$$
$$= - e^{2W} (G_RN)^{[i} (G_L M)^{j]}$$

The field $L$ is not restricted by any equation
and transforms as a compensator under the world--sheet Lorentz
transformations.

\medskip

As for the simplest $D=3$ case, the system of nonlinear
 equations is presented in the form of zero curvature condition
 {\sl and the spectral parameter can be introduced into the associated
 linear system by the trick described for $D=3$ case}
 \footnote{So, in the
 gauge $G_R = I$, $L=W + const$, the form $\Om^{--i}$ becomes
 closed (see Eq.\p{chiral}) }!

\section{ Conclusion and discussion}

We have demonstrated that the zero curvature representation
for the nonlinear equations of motion of bosonic $p$--brane moving in $D$
dimensional space--time originates in  the geometric approach
\cite{lr,barnes,zhelt,bpstv}.
It is given by Maurer--Cartan equation \p{MC0} (or \p{MCs}) involving
\p{Cfv0} (\p{Cfsg}) independent forms  $\Om^{ij}$ and
the forms $\Om^{ab}$,
$\Om^{ai}$ expressed in terms of the world sheet vielbein and the second
fundamental form matrix by the solution \p{eome} of Eq.\p{tor} and
by Eq.\p{2ffi0} respectively.
This representation is
manifestly Lorentz invariant. It is vailed
for any $p$--brane in $D$--dimensional space--time.

The considered zero curvature representation
is based on the map of the $p$--brane world volume into
the coset
$SO(1,D-1)/(SO(1,p) \otimes SO(D-p-1))$ in distinction with Hoppe's
construction \cite{hoppe95} based on the map of $(D-2)$--brane world
volume into the Euclidean $(D-1)$--dimensional hyperplane of the $D$--
dimensional flat target space.

So, the $D$-- dimensional
bosonic $p$--branes shall supplement the list of dynamical systems,
whose equations of motion  can be naturally presented in the form of zero
curvature conditions. Let us remind, that such list includes, in addition
to exactly solvable two--dimensional models, the self-- dual Yang--Mills
system \cite{ward} as well as the full $N=3$ \cite{devog} and $N=2$
\cite{zup} super--Yang--Mills theories.
(See also Refs. \cite{violete,dev})

It is important to investigate the possibility to
introduce a spectral parameter into a linear system \p{aslin} associated
with $p$-- brane equations of motion.
As it was noted in \cite{hoppe95},
this should result, in particular, in the
possibility to obtain infinitely many
conserved currents.

If this
is impossible for general case,
the problem
can be reformulated as one of classification of all
particular cases of $p$-- brane motions in $D$-- dimensional space
time, where an infinite number of conserved currents exists.

As a first step we have investigated here the relation
between the spectral parameter
and the natural $SO(1,1)$ gauge invariance
of the zero curvature representation for the case of string
theory in $D\geq 3$.  We have presented the prescription of the
deformation of the $SO(1,D-1)$ connections, which do not break the zero
curvature condition, and have demonstrated that then the spectral parameter
appears in the associated linear system
as a parameter of constant  $SO(1,1)$ gauge
transformations.

For $D=3$ case such associated linear system \p{lin},\p{deform},
\p{om0h}, \p{om++h}, \p{om--h}
reproduces the Lax matrices \p{Omf}
for nonlinear Liouville equations obtained in Ref. \cite{kulish}
        as well
as one \p{faddeev} obtained in Ref. \cite{faddeev}
            for an appropriate choice of the
$SO(1,1)$ gauge and the values of the constants involved.

\medskip

The form of the geometric approach used here is most suitable for
(doubly) supersymmetric generalization, which was performed in
\cite{bpstv} (see also \cite{bsv}).
It seems to be of interest to construct a zero curvature
representation and associated linear system for
equations of motion of supersymmetric extended
object. Now such supersymmetric generalization
has been built for the simplest cases of $D=3,~N=1~and~2$ superstrings.
This leads to a new version of $n=(1,0)$ and $n=(1,1)$ supersymmetric
Liouville equations
\cite{bsv1}. It can be  shown that the nontrivial dependence on the
spectral parameters can be introduced into the linear systems associated
with these supersymmetric Liouville models by the above prescription.

\medskip

\section {Acknowledgments}
 Author is greatly obliged to  Dmitrij  Sorokin
and  Dmitrij V. Volkov for
significant considerations and illuminating
comments.
Author are thankful to L.D. Faddeev, E.A. Ivanov,
S.O. Krivonos, M. Marvan, V. Nesterenko, L.A. Pastur,
K.S.  Stelle, M. Tonin, F. Topan, M.A. Vasiliev, B.M.
Zupnik and especially to Alexandr A. Zheltukhin   for the interest to
this work and stimulating discussions.  Author would like to thank Prof.
 J.  Lukierski for hospitality at the Institute for Theoretical Physics of
 Wroclaw University, Profs. M.  Tonin and  P.  Pasti for
 hospitality at Padova University and Padova section of INFN, where the
 parts of this work were done.

\medskip

This work was
supported  in  part  by  the
International Science Foundation under the Grant N {\bf RY 9200},
by the INTAS grant {\bf 93 -- 0633}, INTAS and Dutch
Government grant {\bf 94-- 2317}.

\end{document}